\documentclass[letter,longauth,traditabstract]{aa} 

\usepackage{graphicx}
\usepackage{natbib}
\usepackage{txfonts}

\begin{document}

\title{Detection of very-high energy $\gamma$-ray emission from \mbox{\object{NGC 1275}} by the MAGIC telescopes}
\author{
 J.~Aleksi\'c\inst{1} \and
 E.~A.~Alvarez\inst{2} \and
 L.~A.~Antonelli\inst{3} \and
 P.~Antoranz\inst{4} \and
 M.~Asensio\inst{2} \and
 M.~Backes\inst{5} \and
 U.~Barres de Almeida\inst{6} \and
 J.~A.~Barrio\inst{2} \and
 D.~Bastieri\inst{7} \and
 J.~Becerra Gonz\'alez\inst{8,}\inst{9}
% ...
}
\institute { IFAE, Edifici Cn., Campus UAB, E-08193 Bellaterra, Spain
 \and Universidad Complutense, E-28040 Madrid, Spain
 \and INAF National Institute for Astrophysics, I-00136 Rome, Italy
 \and Universit\`a  di Siena, and INFN Pisa, I-53100 Siena, Italy
 \and Technische Universit\"at Dortmund, D-44221 Dortmund, Germany
 \and Max-Planck-Institut f\"ur Physik, D-80805 M\"unchen, Germany
 \and Universit\`a di Padova and INFN, I-35131 Padova, Italy
 \and Inst. de Astrof\'{\i}sica de Canarias, E-38200 La Laguna, Tenerife, Spain
 \and Depto. de Astrof\'{\i}sica, Universidad de La Laguna, E-38206 La Laguna, Spain
% ...
}

\date{Received: 16 December 2011~/~Accepted: 23 January 2012}

\abstract
{
We report on the detection of very-high energy (VHE, $E>100$~GeV) $\gamma$-ray 
emission from \mbox{\object{NGC 1275}}, the central radio galaxy of the Perseus cluster of 
galaxies. The source has been detected by the MAGIC telescopes with a
statistical significance of $6.6~\sigma$ above $100$~GeV in $46$~hr of 
stereo observations carried out between August 2010 and February 2011. 
The measured differential energy spectrum between $70$~GeV and $500$~GeV 
can be described by a power law with a steep spectral index of 
$\Gamma=-4.1\pm0.7_{stat}\pm0.3_{syst}$, and the average flux above $100$~GeV is 
$F_{\gamma}=(1.3\pm0.2_{stat}\pm0.3_{syst})\times10^{-11}~\mathrm{cm^{-2}~s^{-1}}$. 
These results, combined with the power-law spectrum measured in the first 
two years of observations by the \emph{Fermi}--LAT above $100$~MeV, 
with a spectral index of $\Gamma \simeq -2.1$, strongly suggest
the presence of a break or cut-off around tens of GeV in the \mbox{\object{NGC 1275}}
spectrum. The light curve of the source above $100$~GeV does not show 
hints of variability on a month time scale. Finally, we report on the nondetection 
in the present data of the radio galaxy \mbox{\object{IC 310}}, previously discovered by the \emph{Fermi}--LAT and MAGIC.
The derived flux upper limit $F^{U.L.}_{\gamma} (>300~\mathrm{GeV})=1.2\times10^{-12}~\mathrm{cm^{-2}~s^{-1}}$
is a factor $\sim 3$ lower than the mean flux measured by MAGIC between October~2009 and February~2010, 
thus confirming the year time-scale variability of the source at VHE.
}


\titlerunning{Detection of \mbox{\object{NGC 1275}} at VHE by the MAGIC telescopes}

\authorrunning{Aleksi\'c et~al.}

\maketitle

\section{Introduction}
\label{sec:section1}
\mbox{\object{NGC 1275}} ($z=0.0179$), the central dominant galaxy of the Perseus cluster, harbors 
one of the closest active galactic nuclei (AGN), already included in the 
original Seyfert list~\citep{se43}. The AGN is a very bright 
radio source showing an extended jet with Fanaroff-Riley~I morphology 
(e.g.~\citealp{ve94,bu10}). The optical 
emission of the nucleus is variable and strongly polarized from 3$\%$ to 
6$\%$~\citep{ma79,ma83}, implying that the relativistic jet contributes 
significantly to the optical continuum ~\citep{an80}. 
\begingroup
\let\thefootnote\relax\footnotetext{* Corresponding authors: 
S. Lombardi (saverio.lombar\-di@pd.infn.it), P. Colin (colin@mppmu.mpg.de), 
D. Hildebrand (dorothee.hildebrand@phys.ethz.ch), and F. Zandanel (fabio@iaa.es)}
\endgroup
The source has also been classified as a BL~Lac object~\citep{ver78}. 
However, the jet increases its inclination from $10^\circ$ to $20^\circ$
on milliarcsecond scales up to $40^\circ$ to $60^\circ$ at arcsecond scales~\citep{kr92}.
Due to its brightness and proximity this source 
is ideally suited to study the physics of relativistic outflows 
and the ``feedback'' effects of the jet on the cluster environment (e.g.~\citealp{fa08,gal09}).

In fact, \mbox{\object{NGC 1275}} is one of the closest $\gamma$-ray emitting AGN.
It was first unambiguously detected in the high-energy 
(HE, $100$~MeV~$<E<$~$100$~GeV) $\gamma$-ray range by the \emph{Fermi} Large Area 
Telescope (LAT)~\citep{Fermi_1}, during the first four months 
of \emph{all-sky-survey} observations, with an average flux above $100$~MeV of 
$F_{\gamma}=(2.10\pm0.23)\times10^{-7}~\mathrm{cm^{-2}~s^{-1}}$.
The differential energy spectrum  between $100$~MeV and $25$~GeV
was described well by a power law with a spectral index of $\Gamma$~=~$-2.17\pm0.05$.
While no variability was observed during these four months of observations,
subsequent results based on the first year of \emph{Fermi}--LAT observations~\citep{Fermi_2} 
show evidence of flux variability on time scales of months.
Furthermore, the average $\gamma$-ray spectrum show a significant 
deviation from a simple power law, indicating an exponential cut-off 
at the break photon energy of $E_{c}$~=~$(42.2\pm19.6)$~GeV.

More recently, the results obtained from the first two years of \emph{Fermi}--LAT 
observations~\citep{Fermi_3} have given clear evidence for variability on time scales 
of days above $800$~MeV, revealing that several major flaring events occurred during 
the two-year observation period. 
A harder-when-brighter correlation between flux and spectral index was also found.
Brighter and therefore harder $>$~GeV states are then promising for triggering 
observations at very high energy (VHE, $E>100$~GeV). Finally, present upper limits at VHE provided
by MAGIC-I~\citep{al10a} and VERITAS~\citep{acc09} combined with the \emph{Fermi}--LAT 
results mentioned above suggested that \mbox{\object{NGC 1275}} may have a break or cut-off in the 
spectrum around tens of GeV.

The Perseus galaxy cluster contains another $\gamma$-ray source, the 
radio galaxy \mbox{\object{IC 310}}, located at $\sim 0.6^{\circ}$ from \mbox{\object{NGC 1275}}. 
It was discovered in 2010 by the \emph{Fermi}--LAT in the $30$~GeV~--~$300$~GeV energy range~\citep{nsv10} 
and for energies $>260$~GeV by MAGIC~\citep{al10b}.
The combined MAGIC and \emph{Fermi}--LAT spectrum is consistent with a flat spectral energy 
distribution (SED) stretching without a break over more than three orders of magnitude 
in energy ($2$~GeV~--~$7$~TeV). The spectrum at VHE measured by MAGIC has a spectral 
index of $\Gamma$~=~$-2.00\pm0.14$, and the mean flux above 
$300$~GeV, from October 2009 to February 2010, was 
$F_{\gamma}=(3.1\pm0.5)\times10^{-12}~\mathrm{cm^{-2}~s^{-1}}$. 
Strong hints of a week to a year time-scale variability were seen in the MAGIC data.

In this letter we present the results of the observations of \mbox{\object{NGC 1275}}
at VHE performed with the MAGIC telescopes between August 2010 and February 
2011, which resulted in the first detection of the source above $100$~GeV.
The same observational campaign also provided results on the variability at VHE of \mbox{\object{IC 310}}. 
This letter is accompanied by a separate paper dedicated to the study of the Perseus 
cluster environment, focusing on possible VHE $\gamma$-ray emission induced by cosmic~rays~\citep{al11b}. 
The multiwavelength emission of \mbox{\object{NGC 1275}} from radio to VHE will be addressed
in future work.

\section{Observations and analysis}
\label{sec:section2}
The MAGIC system consists of two 17 m dish Imaging Air Cherenkov 
Telescopes (IACTs) located at the Roque de los Muchachos observatory, 
in the Canary Island of La Palma ($28.8^\circ$N, $17.8^\circ$W, 2200~m 
a.s.l.). Since late 2009 the telescopes are working  in 
stereoscopic mode providing an excellent sensitivity of $<0.8\%$ 
of the Crab flux (C.U.)\footnote{In this letter C.U. stands for Crab 
units, defined as the fraction of the Crab Nebula flux given in Eq.~1 
of~\citet{al11a}, which corresponds for energies above $100$~GeV
to $5.4\times10^{-10}~\mathrm{cm^{-2}~s^{-1}}$.} for energies above 
$\sim 300$~GeV in $50$~hr of observations~\citep{al11a} and a trigger
energy threshold of $50$~GeV, which is the lowest among the existing 
IACTs. The MAGIC telescopes are currently 
the most sensitive instrument between $50$~GeV and $200$~GeV, allowing us
to extend up to the TeV scale the observations carried out by the \emph{Fermi}--LAT.

The Perseus galaxy cluster region was observed by the MAGIC telescopes
during two campaigns. The first one was carried out between October 
2009 and February 2010, for a total observation time of $45.3$~hr. This 
survey resulted in the discovery of the radio galaxy \mbox{\object{IC 310}} as VHE emitter~\citep{al10b}. 
The latest campaign (total observation time of $53.6$~hr), 
which resulted in the detection of \mbox{\object{NGC 1275}} at VHE presented in this letter, 
was performed between August 2010 and February 2011.
The source was observed in the wobble mode~\citep{fo94}, 
with data equally split in four pointing positions located symmetrically at 
$0.4^\circ$ from \mbox{\object{NGC 1275}}, in order to ensure optimum sky coverage and background 
estimation. The survey was carried out during dark time at low zenith angles 
(from $12^\circ$ to $36^\circ$), which guaranteed the lowest energy threshold ($\sim 50$~GeV).

The data analysis was performed using the standard MAGIC software package~\citep{al08,aliu09}, 
taking advantage of newly developed stereoscopic analysis 
routines~\citep{mor09,al11a,lom11}. 
The analysis cuts applied to \mbox{\object{NGC 1275}} data were optimized by means of contemporaneous 
Crab Nebula data and Monte Carlo (MC) simulations.

After the application of standard quality checks, $7.9$~hr of data were 
rejected mainly due to nonoptimal atmospheric conditions. The selected 
sample used for deriving the results presented here is therefore composed 
by $45.7$~hr of good quality stereo data.

\section{Results}
\label{sec:section3}
\begin{figure}
\centering
\includegraphics[width=0.45\textwidth]{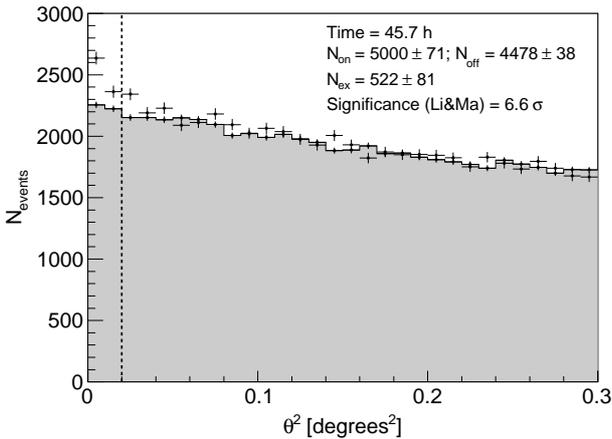}
\caption{$\theta^2$ distributions of the \mbox{\object{NGC 1275}} signal and the background 
estimation from $45.7$~hr of MAGIC stereo observations taken between 
August 2010 and February 2011, above an energy threshold of $100$~GeV.
The region between zero and the vertical dashed line (at $0.02$~degrees$^2$)
represents the signal region.}
\label{fig:figure1}
\end{figure}

The $\theta^2$ distributions\footnote{The $\theta^2$ is the squared angular 
distance between the arrival direction of the events and a given nominal
position (e.g.~\citealp{daum97}).} with respect the signal region
and the background (estimated from 3 distinct 
regions), for energies above $100$~GeV, are shown in Fig.~\ref{fig:figure1}. 
We found an excess of $522\pm81$ events, corresponding to a significance 
of $6.6$ standard deviations ($\sigma$), calculated according to the Eq.~17 
of~\cite{lm83}. It is worth noting that the background estimation
is not affected by a possible \mbox{\object{IC 310}} $\gamma$-ray contribution, since the latter
source was not detected in the present data.

The \mbox{\object{NGC 1275}} differential energy spectrum measured by MAGIC between $70$~GeV 
and $500$~GeV can be described by a simple power law ($\chi^{2}/n_{dof}=0.76/1$)
\begin{equation}
\frac{\mbox{d}N}{\mbox{d}E} = (3.1\pm1.0_{stat}\pm0.7_{syst}) \times 10^{-10} \left(\frac{E}{\mathrm{100~GeV}}\right)^{\Gamma},
\end{equation}
in units of $\mathrm{cm^{-2}~s^{-1}~TeV^{-1}}$, 
with $\Gamma=-4.1\pm0.7_{stat}\pm0.3_{syst}$\footnote{
The systematic errors of the flux normalization and the energy spectral slope
considered here have been estimated to be $23\%$ and $\pm0.3$, respectively, whereas
the systematic error on the energy scale is $17\%$. These values 
are more conservative than those presented in~\cite{al11a}, given the flux weakness and the
spectral steepness of \mbox{\object{NGC 1275}}, as measured by MAGIC.}.
The mean flux above $100$~GeV is 
$F_{\gamma}=(1.3\pm0.2_{stat}$~$\pm$~$0.3_{syst})\times10^{-11}~\mathrm{cm^{-2}~s^{-1}}$,
corresponding to $(2.5\pm0.4_{stat}\pm0.6_{syst})\%$~C.U. The steepness of the spectral index measured 
by MAGIC strongly supports the presence of a break or cut-off in the \mbox{\object{NGC 1275}} spectrum 
around tens of GeV, as already suggested by the \emph{Fermi}-–LAT results~\citep{Fermi_2,Fermi_3},
and is consistent with the upper limits on the flux at VHE provided by MAGIC-I~\citep{al10a} and 
VERITAS~\citep{acc09}. The rapid decline in the spectrum, which causes the \mbox{\object{NGC 1275}}
signal to vanish above approximately $500$~GeV, permits investigating possible VHE 
$\gamma$-ray emissions induced by cosmic~rays in the Perseus cluster environment above that 
energy~\citep{al11b}.

\begin{figure}
\centering
\includegraphics[width=0.45\textwidth]{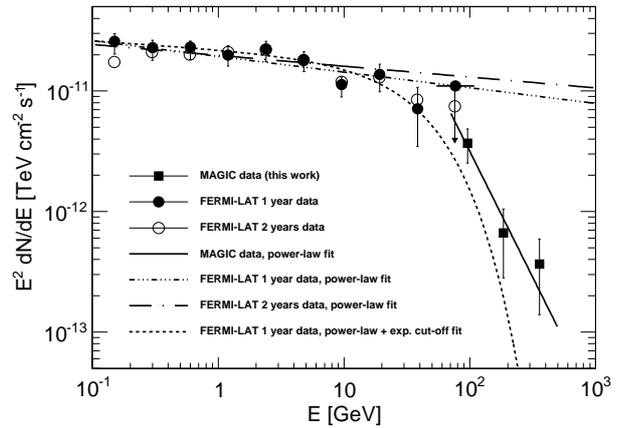}
\caption{\mbox{\object{NGC 1275}} spectral energy distribution measured by MAGIC
between $70$~GeV and $500$~GeV (filled squares), together with the 
results above $100$~MeV achieved from the first year (filled circles,~\citealp{Fermi_2}), 
and from the first two years (open circles,~\citealp{Fermi_3}) of the \emph{Fermi}--LAT \emph{all-sky-survey} observations. 
The power-law fits to the \emph{Fermi}--LAT data (extrapolated up to $1$~TeV) 
are also shown, together with the exponential power-law fit provided in~\cite{Fermi_2}.}
\label{fig:figure2}
\end{figure}

In Fig.~\ref{fig:figure2}, the SED measured by MAGIC is compared with the results in the 
$100$~MeV~--~$100$~GeV range provided by the \emph{Fermi}--LAT, averaging \emph{Fermi} data over the 
first year~\citep{Fermi_2} and the first two years~\citep{Fermi_3}. The comparison suggests that a significant 
spectral steepening occurs around $\sim 100$~GeV. However, the present non-simultaneous 
data do not allow discussing whether the spectral change corresponds to a break between 
two power laws or exponential cut-off. 

\begin{figure}
\centering
\includegraphics[width=0.45\textwidth]{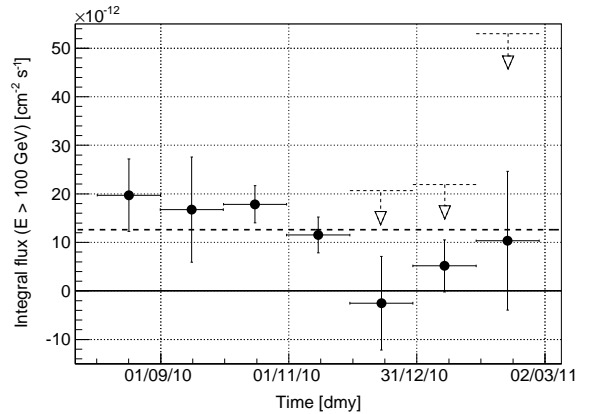}
\caption{\mbox{\object{NGC 1275}} light curve between August 2010 and February 2011 above an energy
threshold of $100$~GeV, and with a month time-scale binning. No hints of variability 
are seen in the data. The dashed horizontal line represents the constant function resulting from the fit to the data.
For the December 2010, January 2011, and February 2011 data, the upper limits on the flux above $100$~GeV 
for a spectral index of $\Gamma=-4.0$ (calculated using the~\citealp{rol05} method 
with a confidence level of $95\%$, and a total systematic uncertainty of $30\%$) are also shown (open dashed arrows).}
\label{fig:figure3}
\end{figure}

The August 2010 to February 2011 light curve of \mbox{\object{NGC 1275}} 
computed for an energy threshold of $100$~GeV and with a monthly binning is shown in Fig.~\ref{fig:figure3}. 
No evidence of variability can be derived from these measurements.
In fact, fitting the light curve with a constant flux hypothesis yields a $\chi^{2}/n_{dof}=7.4/6$, 
corresponding to a probability $P(\chi^2)=0.29$.

\begin{figure}
\centering
\includegraphics[width=0.45\textwidth]{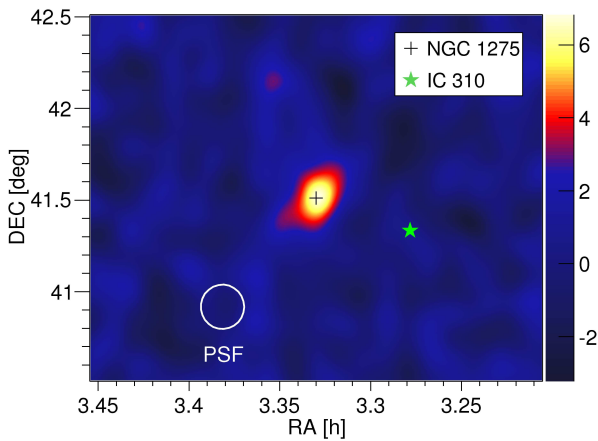}
\caption{Significance skymap of the central region of the Perseus galaxy 
cluster from $45.7$~hr of MAGIC stereo observations taken between August 
2010 and February 2011, above an energy threshold of $100$~GeV. The \mbox{\object{NGC 1275}} 
position is marked with a black cross, whereas the position of the radio 
galaxy \mbox{\object{IC 310}} is shown by a green star. The PSF of about $0.12^\circ$ 
is also displayed. The hot spot in the map at a significance level $>6~\sigma$ 
is consistent with a point-like emission coming from the \mbox{\object{NGC 1275}} sky position.
}
\label{fig:figure4}
\end{figure}

The significance skymap of the central region of the Perseus cluster above $100$~GeV 
is shown in Fig.~\ref{fig:figure4}. A hot spot at $>6~\sigma$ significance level consistent 
with the sky position of \mbox{\object{NGC 1275}} is present. The position of the radio galaxy \mbox{\object{IC 310}} 
is also shown. No significant excess events coming from 
\mbox{\object{IC 310}} have been found in the observations presented here. The corresponding integral
flux upper limit above $300$~GeV (performed using the~\citealp{rol05} method,
with a confidence level of $95\%$, and a total systematic uncertainty of $30\%$) 
is $F^{U.L.}_{\gamma}(>300~\mathrm{GeV})$~=~$1.2\times10^{-12}~\mathrm{cm^{-2}~s^{-1}}$,
for a spectral index of $\Gamma$~=~$-2.0$ (i.e. the spectral index of the source previously
measured by MAGIC). This value is about a factor 3 lower than the average integral flux 
$F_{\gamma} (>300~\mathrm{GeV})=(3.1\pm0.5)\times10^{-12}~\mathrm{cm^{-2}~s^{-1}}$
measured by MAGIC from October 2009 to February 2010~\citep{al10b},
thereby confirming the variability of the latter source on a year's time scale.

\section{Conclusions}
\label{sec:section4}
The MAGIC telescopes have detected VHE $\gamma$-ray emission from \mbox{\object{NGC 1275}},
the central radio galaxy in the Perseus cluster, at a statistical significance 
of $6.6~\sigma$ from observations performed between August 2010 and February 2011.
The corresponding average flux above $100$~GeV is 
$F_{\gamma}=(1.3\pm0.2_{stat}\pm0.3_{syst})\times10^{-11}~\mathrm{cm^{-2}~s^{-1}}$. %$(2.5\pm0.4_{stat})\%$ C.U. 
This is the fourth nearby radio galaxy detected at VHE, after \mbox{\object{M 87}}~\citep{ah03,ah06,al08_M87,acc08}, 
\mbox{\object{Cen A}}~\citep{ah09}, and \mbox{\object{IC 310}}~\citep{al10b}. 
The MAGIC observation yields a spectrum that can be fitted between $70$~GeV 
and $500$~GeV by a simple power law with a spectral index of 
$\Gamma=-4.1\pm0.7_{stat}\pm0.3_{syst}$. This result, combined with previous 
\emph{Fermi}-–LAT results~\citep{Fermi_1,Fermi_2,Fermi_3}, 
showing a power-law spectrum with a spectral index of $\Gamma \simeq -2.1$ 
above $100$~MeV, strongly suggests the presence of a break or cut-off
around tens of GeV in the \mbox{\object{NGC 1275}} spectrum. No evidence of variability on 
month time scale has been found above $100$~GeV. Finally, the variability on a year 
time scale of the source \mbox{\object{IC 310}}~\citep{nsv10,al10b} has been 
confirmed. The upper limit above $300$~GeV presented here is in fact about a factor 3 
lower than the flux measured by MAGIC between October 2009 and February 2010.

\begin{acknowledgements}
We would like to thank the Instituto de Astrof\'{\i}sica de
Canarias for the excellent working conditions at the
Observatorio del Roque de los Muchachos in La Palma.
The support of the German BMBF and MPG, the Italian INFN,
the Swiss National Fund SNF, and the Spanish MICINN is
gratefully acknowledged. This work was also supported by
the Marie Curie program, by the CPAN CSD2007-00042 and MultiDark
CSD2009-00064 projects of the Spanish Consolider-Ingenio 2010
program, by grant DO02-353 of the Bulgarian NSF, by grant 127740 of
the Academy of Finland, by the YIP of the Helmholtz Gemeinschaft,
by the DFG Cluster of Excellence ``Origin and Structure of the
Universe'', by the DFG Collaborative Research Centers SFB823/C4 and SFB876/C3,
and by the Polish MNiSzW grant 745/N-HESS-MAGIC/2010/0.
C.P. gratefully acknowledges financial support of the Klaus
Tschira Foundation.
\end{acknowledgements}

\end{document}